\documentstyle[twoside,psfig]{article}

\catcode`\@=11
\long\def\@makefntext#1{
\protect\noindent \hbox to 3.2pt {\hskip-.9pt  
$^{{\eightrm\@thefnmark}}$\hfil}#1\hfill}               

\def\@makefnmark{\hbox to 0pt{$^{\@thefnmark}$\hss}}    
        
\def\ps@myheadings{\let\@mkboth\@gobbletwo
\def\@oddhead{\hbox{}
\rightmark\hfil\eightrm\thepage}   
\def\@oddfoot{}\def\@evenhead{\eightrm\thepage\hfil
\leftmark\hbox{}}\def\@evenfoot{}
\def\sectionmark##1{}\def\subsectionmark##1{}}

\oddsidemargin=\evensidemargin
\newcounter{sectionc}\newcounter{subsectionc}\newcounter{subsubsectionc}
\renewcommand{\section}[1] {\vspace{12pt}\addtocounter{sectionc}{1} 
\setcounter{subsectionc}{0}\setcounter{subsubsectionc}{0}\noindent 
        {\tenbf\thesectionc. #1}\par\vspace{5pt}}
\renewcommand{\subsection}[1] {\vspace{12pt}\addtocounter{subsectionc}{1} 
        \setcounter{subsubsectionc}{0}\noindent 
        {\bf\thesectionc.\thesubsectionc. {\kern1pt \bfit #1}}\par\vspace{5pt}}
\renewcommand{\subsubsection}[1] {\vspace{12pt}\addtocounter{subsubsectionc}{1}
        \noindent{\tenrm\thesectionc.\thesubsectionc.\thesubsubsectionc.
        {\kern1pt \tenit #1}}\par\vspace{5pt}}
\newcommand{\nonumsection}[1] {\vspace{12pt}\noindent{\tenbf #1}
        \par\vspace{5pt}}

\newcounter{appendixc}
\newcounter{subappendixc}[appendixc]
\newcounter{subsubappendixc}[subappendixc]
\renewcommand{\thesubappendixc}{\Alph{appendixc}.\arabic{subappendixc}}
\renewcommand{\thesubsubappendixc}
        {\Alph{appendixc}.\arabic{subappendixc}.\arabic{subsubappendixc}}

\renewcommand{\appendix}[1] {\vspace{12pt}
        \refstepcounter{appendixc}
        \setcounter{figure}{0}
        \setcounter{table}{0}
        \setcounter{lemma}{0}
        \setcounter{theorem}{0}
        \setcounter{corollary}{0}
        \setcounter{definition}{0}
        \setcounter{equation}{0}
        \renewcommand{\thefigure}{\Alph{apA dynamical model of a GRID marketpendixc}.\arabic{figure}}
        \renewcommand{\thetable}{\Alph{appendixc}.\arabic{table}}
        \renewcommand{\theappendixc}{\Alph{appendixc}}
        \renewcommand{\thelemma}{\Alph{appendixc}.\arabic{lemma}}
        \renewcommand{\thetheorem}{\Alph{appendixc}.\arabic{theorem}}
        \renewcommand{\thedefinition}{\Alph{appendixc}.\arabic{definition}}
        \renewcommand{\thecorollary}{\Alph{appendixc}.\arabic{corollary}}
        \noindent{\tenbf Appendix \theappendixc #1}\par\vspace{5pt}}
\newcommand{\subappendix}[1] {\vspace{12pt}
        \refstepcounter{subappendixc}
        \noindent{\bf Appendix \thesubappendixc. {\kern1pt \bfit #1}}
        \par\vspace{5pt}}
\newcommand{\subsubappendix}[1] {\vspace{12pt}
        \refstepcounter{subsubappendixc}
        \noindent{\rm Appendix \thesubsubappendixc. {\kern1pt \tenit #1}}
        \par\vspace{5pt}}

\newcommand{\textlineskip}{\baselineskip=13pt}
\newcommand{\smalllineskip}{\baselineskip=10pt}

\def\eightcirc{
\begin{picture}(0,0)
\put(4.4,1.8){\circle{6.5}}
\end{picture}}
\def\eightcopyright{\eightcirc\kern2.7pt\hbox{\eightrm c}}

\newcommand{\copyrightheading}[1]
        {\vspace*{-2.5cm}\smalllineskip{\flushleft
        {\footnotesize International Journal of Modern Physics C #1}\\
        {\footnotesize $\eightcopyright$\, World Scientific Publishing
         Company}\\
         }}

\def\abstracts#1#2#3{{
        \centering{\begin{minipage}{4.5in}\footnotesize\baselineskip=10pt
        \parindent=0pt #1\par 
        \parindent=15pt #2\par
        \parindent=15pt #3
        \end{minipage}}\par}} 

\def\keywords#1{{
        \centering{\begin{minipage}{4.5in}\footnotesize\baselineskip=10pt
        {\footnotesize\it Keywords}\/: #1
        \end{minipage}}\par}}

\renewenvironment{thebibliography}[1]
        {\frenchspacing
         \ninerm\baselineskip=11pt
         \begin{list}{\arabic{enumi}.}
        {\usecounter{enumi}\setlength{\parsep}{0pt}     
         \setlength{\leftmargin 12.7pt}{\rightmargin 0pt} 
         \setlength{\itemsep}{0pt} \settowidth
        {\labelwidth}{#1.}\sloppy}}{\end{list}}

\newcounter{itemlistc}
\newcounter{romanlistc}
\newcounter{alphlistc}
\newcounter{arabiclistc}

\newcommand{\fcaption}[1]{
        \refstepcounter{figure}
        \setbox\@tempboxa = \hbox{\footnotesize Fig.~\thefigure. #1}
        \ifdim \wd\@tempboxa > 5in
           {\begin{center}
        \parbox{5in}{\footnotesize\smalllineskip Fig.~\thefigure. #1}
            \end{center}}
        \else
             {\begin{center}
             {\footnotesize Fig.~\thefigure. #1}
              \end{center}}
        \fi}

\newcommand{\tcaption}[1]{
        \refstepcounter{table}
        \setbox\@tempboxa = \hbox{\footnotesize Table~\thetable. #1}
        \ifdim \wd\@tempboxa > 5in
           {\begin{center}
        \parbox{5in}{\footnotesize\smalllineskip Table~\thetable. #1}
            \end{center}}
        \else
             {\begin{center}
             {\footnotesize Table~\thetable. #1}
              \end{center}}
        \fi}

\def\@citex[#1]#2{\if@filesw\immediate\write\@auxout
        {\string\citation{#2}}\fi
\def\@citea{}\@cite{\@for\@citeb:=#2\do
        {\@citea\def\@citea{,}\@ifundefined
        {b@\@citeb}{{\bf ?}\@warning
        {Citation `\@citeb' on page \thepage \space undefined}}
        {\csname b@\@citeb\endcsname}}}{#1}}

\newif\if@cghi
\def\cite{\@cghitrue\@ifnextchar [{\@tempswatrue
        \@citex}{\@tempswafalse\@citex[]}}
\def\citelow{\@cghifalse\@ifnextchar [{\@tempswatrue
        \@citex}{\@tempswafalse\@citex[]}}
\def\@cite#1#2{{$\null^{#1}$\if@tempswa\typeout
        {IJCGA warning: optional citation argument 
        ignored: `#2'} \fi}}

\def\pmb#1{\setbox0=\hbox{#1}
        \kern-.025em\copy0\kern-\wd0
        \kern.05em\copy0\kern-\wd0
        \kern-.025em\raise.0433em\box0}

\def\fnt#1#2{\footnotetext{\kern-.3em
        {$^{\mbox{\scriptsize #1}}$}{#2}}}

\def\ps@myheadings{%
    \let\@oddfoot\@empty\let\@evenfoot\@empty
    \def\@evenhead{\slshape\leftmark\hfil}
    \def\@oddhead{\hfil{\slshape\rightmark}}
    \let\@mkboth\@gobbletwo
    \let\sectionmark\@gobble
    \let\subsectionmark\@gobble
    }
\font\tenrm=cmr10
\font\tenit=cmti10 
\font\tenbf=cmbx10
\font\bfit=cmbxti10 at 10pt
\font\ninerm=cmr9

\font\eightrm=cmr8

\def\qed{\hbox{${\vcenter{\vbox{                    
   \hrule height 0.4pt\hbox{\vrule width 0.4pt height 6pt
   \kern5pt\vrule width 0.4pt}\hrule height 0.4pt}}}$}}

\def\bsc{{\sc a\kern-6.4pt\sc a\kern-6.4pt\sc a}}       
\def\bflatex{\bf L\kern-.30em\raise.3ex\hbox{\bsc}\kern-.14em 
T\kern-.1667em\lower.7ex\hbox{E}\kern-.125em X} 

\begin{document}
\setlength{\textheight}{7.7truein}  

\thispagestyle{empty}

\normalsize\textlineskip

\setcounter{page}{1}

\copyrightheading{}                     {Vol. 0, No. 0 (2004) 000--000}

\vspace*{0.88truein}

\centerline{\bf Changing Opinions in a Changing World:}
\vspace*{0.035truein}
\centerline{\bf a New Perspective in Sociophysics}
\vspace*{0.37truein}

\centerline{\footnotesize Alessandro Pluchino}
\baselineskip=12pt
\centerline{\footnotesize\it Dipartimento di Fisica e Astronomia and 
Infn sezione di Catania , Universit\'a di  Catania} 
\baselineskip=10pt
\centerline{\footnotesize\it Catania, I-95123 , Italy }
\centerline{\footnotesize\it E-mail: alessandro.pluchino@ct.infn.it}

\vspace*{15pt}          
\centerline{\footnotesize Vito Latora}
\centerline{\footnotesize\it Dipartimento di Fisica e Astronomia, and 
Infn sezione di Catania , Universit\'a di  Catania}
\baselineskip=10pt
\centerline{\footnotesize\it Catania, I-95123 , Italy }
\centerline{\footnotesize\it E-mail: vito.latora@ct.infn.it}

\vspace*{15pt}          
\centerline{\footnotesize Andrea Rapisarda}
\baselineskip=12pt
\centerline{\footnotesize\it Dipartimento di Fisica e Astronomia, and 
Infn sezione di Catania , Universit\'a di  Catania}
\baselineskip=10pt
\centerline{\footnotesize\it Catania, I-95123 , Italy }
\centerline{\footnotesize\it E-mail: andrea.rapisarda@ct.infn.it}

\vspace*{0.225truein}

\vspace*{0.25truein}
\abstracts{We propose a new model of opinion formation,  
the {\it Opinion Changing Rate} (OCR) model. 
Instead of investigating the conditions that allow 
consensus in a world of agents with different opinions, 
we study under which conditions a group of agents with a 
different natural tendency ($rate$) to change opinion can find 
agreement. 
The OCR is a modified version  of the Kuramoto model, 
one of the simplest models for synchronization in biological systems, 
here adapted to a social context. 
By means of several  numerical simulations we illustrate the richness of
the OCR model dynamics and its social implications. 
}{}{}

\vspace*{5pt}


\vspace*{1pt}\textlineskip      
\section{Introduction}          
\vspace*{-0.5pt}
\noindent
The world changes and we change with it.
But  everyone in a different way. There are conservative people
that strongly tend to maintain their opinion or their style 
of life against everything and everyone.
There are more flexible people that change ideas very easily
and follow the current fashions  and trends. Finally,  there are
those who run faster than the rest of the world anticipating
the others. These different tendencies can be interpreted 
as a continuous spectrum of different degrees of natural
inclination to changes. By changes here we mean change of opinions
or more in general change of ideas, habits, style of
life or way of thinking. 
\\
The last  years have seen a large interest in the physics 
community towards the description and modeling of 
social systems\cite{weidlich,staufrev,stauf}. 
In particular, Monte Carlo simulations have become an important part 
of {\it sociophysics},
enlarging the field of interdisciplinary
applications of statistical physics. 
Many of the most popular sociophysics models 
\cite{santoreview} deal with opinion 
dynamics and consensus formation but have the limitation of not
taking into account the individual inclination to change,
a peculiar feature of any social system. In fact, as Charles Darwin wrote, 
``it is not the strongest that survives, nor the most intelligent; 
it is the one that is most adaptable to change''.
\\
In this paper we show how such an individual inclination
to change, differently distributed in a group of people, can affect the opinion dynamics of the group itself.
The first advantage of adopting this new perspective is 
the possibility of using the knowledge accumulated in the 
study of coupled oscillators. 
If we switch from  the question 
{\it "Could agents with initial different opinions reach a final  agreement?"} 
into the more realistic  one  
{\it "Could agents with a  different
 natural tendency to change opinion reach a final agreement?"},
 we introduce the concept of natural
opinion changing rate, that is very similar to the
characteristic frequency of an oscillator. In such a way, 
we can treat consensus as a peculiar kind of synchronization 
(frequency locking) which has been very well studied in different contexts
by means of the Kuramoto model\cite{kuramoto_model}. 
The paper is organized as follows. In Section 2 we briefly 
rewiew the Kuramoto model , a system of ordinary
differential equations introduced 
to model synchronization in biological coupled oscillators. 
In Section 3 we propose the Opinion Changing Rate (OCR) 
model, a new model of opinion dynamics inspired by the Kuramoto model.
We discuss in detail the definitions, and we explore the 
rich dynamics of the model by means of 
extensive molecular dynamics simulations.   
In Section 4 we draw the conclusions and we 
outline possible generalizations and future developments.

\vspace*{1pt}\textlineskip      
\section{The Kuramoto model}            
\vspace*{-0.5pt}
\noindent
Coupled oscillators are ubiquitous in nature \cite{strogatz00,piko}.
A large number of biological, physical and chemical systems
can be thought of as a large ensemble of weakly
interacting nonidentical oscillators.  
Examples include flashing fireflys \cite{buck88}, 
and chorusing crickets \cite{walker69,sismondo90}, 
pacemaker cells in the heart \cite{peskin75} and in the brain, 
coupled laser arrays \cite{piko}.
One interesting phenomenon common to all these systems is 
their ability to collectively synchronize: a large number of the oscillators lock onto a common 
frequence despite the difference in their natural frequencies. 
The most striking visual example is given by the synchronous 
flashing of fireflys observed in summer season in some region 
of South Asia. At night thousands of fireflies meet on the trees 
along the river. Suddenly they start to emit flashes of light.
Initially they act incoherently and after a short time
the all flash in unison, creating one of the most beautiful 
effects of synchronization \cite{sync}.

Researches on synchronization aim at understanding what are
the basic mechanisms responsible for the collective synchronous 
behavior in a given population. 
The Kuramoto model of coupled oscillators is one of the simplest 
and most successful models for synchronization. 
It is simple enough to be analytically solvable, still retaining 
the basic principles to produce a rich variety of dynamical regimes
and synchronization patterns. 
It was originally proposed in 1975 by Kuramoto \cite{kuramoto_model} 
as an analytically tractable version of the
Winfree's mean-field model for large populations of biological 
oscillators \cite{winfree67}. Since then  
the Kuramoto model has been analyzed more deeply and several extensions 
and generalizations have been considered 
\cite{SK86_S88,SMM92,ritort,strogatz00,piko,acebron04}.
It has also been linked to several physical problems,
including Landau damping in plasmas \cite{SMM92} and the dynamics of
Josephson junction arrays \cite{WCS96_98}. 
\\
The Kuramoto model  describes a population of $N$ periodic
oscillators having natural frequencies $\omega_i$ and
coupled through the sine of their phase differences.
The dynamics of the model is given by
\begin{equation}
    \dot{\theta_i} (t)  = \omega_i + \frac{K}{N} \sum_{j=1}^N
      \sin ( \theta_j  - \theta_i )  ~~~~~i=1,\dots,N
\label{kuramoto_eq1}
\end{equation}
where $\theta_i (t)$ is the phase (angle) of the $i$th oscillator at time $t$,
while $\omega_i$ is its intrinsic frequency randomly drawn from
a symmetric, unimodal distribution $g(\omega)$ with a first moment
$\omega_0$ (typically a Gaussian distribution or a uniform one). These
natural frequencies $\omega_i$  are time-independent. 
The sum in the above equation is running over all the
oscillators so that this is an example of a globally coupled system. 
The parameter $K \geq 0$ measures the coupling strength in the global coupling term.
For small  values of $K$, each oscillator tends to run independently with its own
frequency,  while for large values of  $K$,  the coupling tends to synchronize (in phase and frequency)
the oscillator with all the others.
Notice that the equations (\ref{kuramoto_eq1}) can be
transformed into an equivalent system of phase oscillators
with a zero mean frequency, by a suitable transformation
$\theta_i \rightarrow \theta_i - \Omega t$ to a rotating frame with
a velocity $\Omega = \omega_0$.

In a beautiful analysis, Kuramoto showed that the model,
despite the difference in the natural frequencies of the
oscillators, exhibits a
spontaneous transition from incoherence to collective synchronization,
as the coupling strength is increased beyond  a certain threshold $K_c$
\cite{kuramoto_model}.
Kuramoto studied the system in terms of a complex mean field
order parameter defined as
\begin{equation}
r e^{i \Psi} = \frac{1}{N} \sum_{j=1}^N   e^{i \theta_j}~~~,
\label{kuramoto_orderparameter}
\end{equation}
where the magnitude $0 \le r(t) \le 1$ is a measure of the
coherence of the population and $\Psi(t)$ is the average
phase.
The system of equations (\ref{kuramoto_eq1}) can be rewritten, in terms 
of the mean field variable $r$ and $\Psi$, as
\begin{equation}
    \dot{\theta_i} (t)  = \omega_i +
K r   \sin ( \Psi  - \theta_i )  ~~~~~i=1,\dots,N~~.
\label{kuramoto_eq2}
\end{equation}
In the limit of infinitely many oscillators, 
the system can be described by a continuous 
probability density $\rho(\theta,\omega,t)$, satisfying the
normalization condition
\begin{equation}
\int_{0}^{2 \pi} \rho(\theta,\omega,t) ~d \theta = 1
\end{equation}
and the order parameter of  Eq.~(\ref{kuramoto_orderparameter})
can be expressed in terms of  $\rho(\theta,\omega,t)$ as:
\begin{equation}
r e^{i \Psi} = \int_{0}^{2 \pi} \int_{- \infty}^{+ \infty}
e^{i \theta} \rho(\theta,\omega,t)~g(\omega) ~d \theta ~d \omega
\label{kuramoto_orderparameter2}
\end{equation} 
The dynamics of the system
in terms of the probability density is governed by the
continuity equation:
\begin{equation}
 \frac{\partial \rho}{\partial t} + \frac{\partial }{\partial \theta}
\{ [ w + Kr \sin (\Psi - \theta)] \rho \} = 0
\label{continuous}
\end{equation}
since each oscillator moves with an angular velocity (frequency) $v_i = \dot{\theta_i} $ given by Eq.(\ref{kuramoto_eq2}).
\\
The main conclusions that can be drawn from a simple analysis of the 
previous formulas and from the results of numerical simulations are the 
following. 

\begin{itemize}

\item
In the limit $K \rightarrow 0$, equations (\ref{kuramoto_eq2}) give
$\theta_i(t) \approx \omega_i t + \theta_i(0)~ \forall i=1,...,N$,
and consequently the system is in a {\it phase incoherent state}.
The individual phase variables $\theta_i(t)$  are not correlated and
the average of $e^{\theta_i(t)}$ over all oscillators,
the sum on the right hand side of Eq. (\ref{kuramoto_orderparameter}),
is expected to go to zero with the system size as $ N^{-1/2}$,
yielding, for $N\rightarrow \infty$ and $t \rightarrow \infty$, a vanishing saturation value of the order parameter, i.e. $r_\infty \rightarrow 0$.

\item
On the other hand, in the strong coupling limit $K \rightarrow \infty$,
Eq. (\ref{continuous})
has a stationary solution in which
all the oscillators have almost the same phase $\theta_i(t) \approx \Psi(t)~ \forall i$.
Consequentely $r_\infty \approx1$ and we have a {\it global synchronization}. 
It is important to stress, however, that since Eq.(\ref{kuramoto_eq1})
implicates the impossibility of simultaneously 
synchronize both phase and frequency, here  synchronization  means  
that all the oscillators have, for $t \rightarrow \infty$, \textit{exactly} 
the same long term frequency (frequency locking) and \textit{almost} 
the same phase. In fact there is a small phase shift between 
the oscillators and such phase shift is conserved in time 
\cite{strogatz00,sebino}.

\begin{figure}[htbp] 
\vspace*{8pt}
\centerline{\psfig{figure=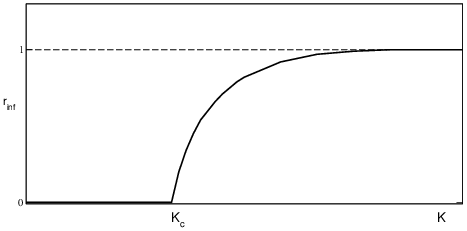,width=11truecm,angle=0}} 
\vspace*{8pt}
\fcaption{Asymptotic order parameter $r_\infty$ as a function of the coupling in the Kuramoto model}
\end{figure}

\item
In the intermediate situation of finite coupling a less
degree of 
synchronization with $0 \le r_\infty \le 1$ is observed (\textit{partial synchronization}).
The stationary solution of Eq.(\ref{continuous}) is a
{\it phase coherent state} in which part of the
oscillators clump around the mean phase $\Psi(t)$,
while the others move freely out of synchrony. In fact,
the typical oscillator is running with angular velocity
$v_i = \omega_i + K r   \sin ( \Psi  - \theta_i )$.
The oscillators in the coherent population will become
stable locked at an angle such that
$\omega_i =  K r   \sin ( \theta_i - \Psi  )$ in the frame of
reference in which the common velocity is set 
equal to zero.  Such a condition
cannot be satisfied by the oscillators with $ | \omega | \ge Kr$: 
these oscillators cannot be locked 
and their stationary density obeys $v \rho =$ constant
according to Eq. (\ref{continuous}).
\\
The appearance of the coherent state happens 
at a critical coupling strength $K_c$.
\end{itemize}

Summing up, as $N$ approaches infinity, the magnitude $r_\infty$
of the complex mean field after a transient  time should be zero in
the incoherent state with $K \le K_c$ and different from zero
in the coherent state with $K > K_c$. 
Actually, as $K$ increases  beyond $K_c$, more and more 
oscillators will be recruited toward the mean phase $\Psi(t)$
and $r_\infty$ is expected to continuously increase from zero
to one (see Fig.1).
\\
The details of this dynamical phase transition 
for the globally coupled periodic oscillators 
have been studied both analytically and numerically.
For instance, a detailed description of the Kuramoto's analytical
arguments to study the bifurcation from the incoherence state can 
be found in Ref.~\cite{strogatz00}. 
Here we only summarize the main results obtained.
The critical value $K_c$ can be expressed as a function of 
$g(\omega)$, namely $K_c= 2/(\pi g(0))$.  
For instance in the simple case of a Lorentzian distribution
$g(\omega) = (\gamma/\pi)/(\gamma^2 + \omega^2)$ it results
$K_c= 2 \gamma$ and it is also possible to obtain an analytical expression for the magnitude
of the order parameter, $r= \sqrt{1- K_c/K}$, with $K> K_c$.

\section{The "Opinion Changing Rate" Model}
\noindent
Usually, consensus models \cite{Sznajd,Deff,HK} in sociophysics deal 
with $N$ individuals (or agents): eeach individual $i=1,...,N$ is   
characterized by an opinion $x_i$ and is in interaction  
with all the others.   
The opinions are integer numbers (for instance +1 or --1) in the 
Sznajd \cite{Sznajd} model, or real numbers in the range [0,1]   
in the Deffuant et al. \cite{Deff} and in the Hegselmann and Krause
\cite{HK} model.
The interaction dynamics of such models is governed by very simple 
deterministic rules. For instance, in the Sznajd model 
a pair of neighbouring agents on a square 
lattice convinces its six neighbours of the pair opinion if and 
only if the two agents of the pair share the same opinion. 
In the Deffuant et al.  model a pair 
of agents $i$ and $k$ is selected at random. 
If the opinions $x_i$ and $x_k$ differ by more than a fixed 
parameter $\epsilon$ nothing happens because the two agents 
think too differently to interact and eventually find an agreement. 
If $| x_i - x_k | < \epsilon$ then both opinions get closer 
to each other by an amount $\mu \cdot |x_i - x_k| $, tuned by the parameter $\mu$.
The main goal of the opinion dynamics models is to figure 
out whether and when a complete or partial consensus can emerge
out of initially different opinions, no matter how long it takes for the
consensus to be reached.
For instance the Deffuant et al. model presents a rich
scenario of outcomes as the two control parameters 
$\epsilon$ and $\mu$  are varied. 
\\
%
\begin{figure}[htbp] 
\vspace*{8pt}
\centerline{\psfig{figure=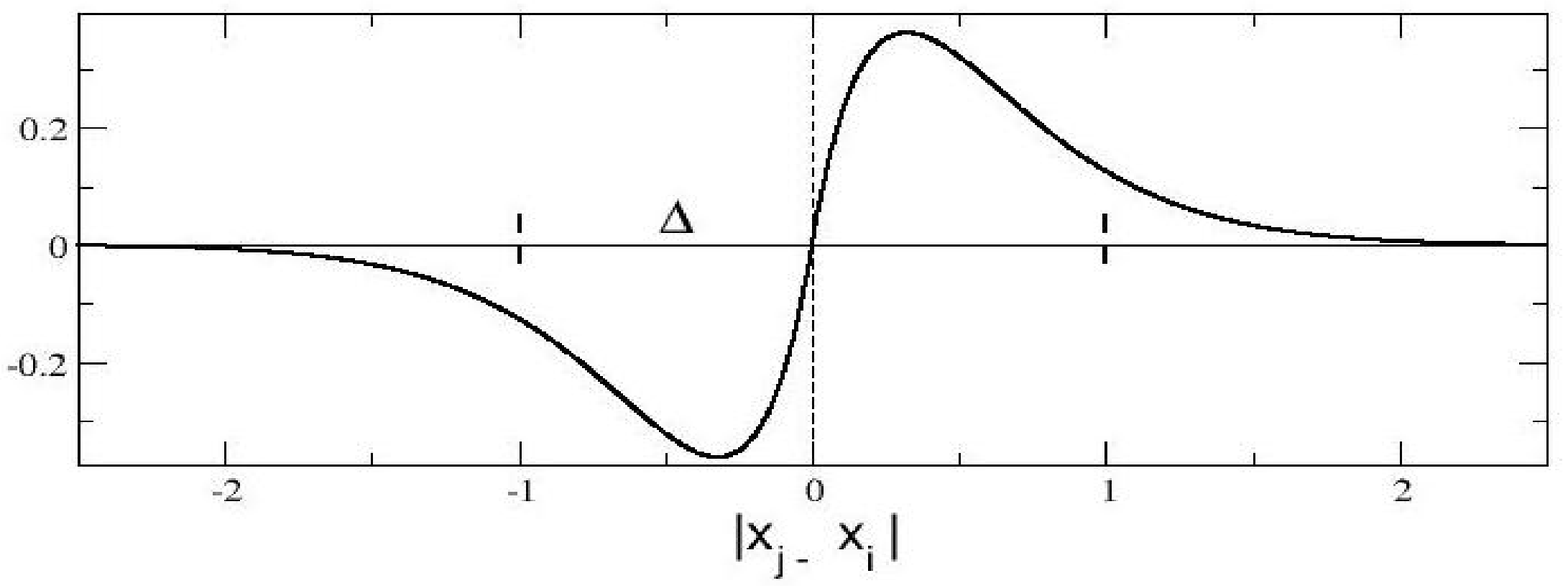,width=12truecm,angle=0}} 
\vspace*{8pt}
\fcaption{Behavior of the coupling term in Eq.(\ref{OCR_eq1}), 
namely $sin ( x_j  - x_i ) e^{- \alpha |x_j  - x_i|}$, 
as a function of the reciprocal distance between two opinions. 
We set $\alpha=3$ and we consider a distribution of the
initial individual opinions $x_i(t=0)$ in the range 
[$-\Delta,\Delta$], with $\Delta=1$, in order to ensure that, 
at the beginning of the simulation, the interaction between each 
couple of agents is not negligeable.}
\label{fig1bis}
\end{figure}
%
The first difference between such models and our approach 
is that we are interested mainly 
on the dynamical aspects of the consensus formation and not only on the equilibrium ones.
Of course, at variance with the phases in the Kuramoto model, in our model
 we do not  want periodic opinions  nor limited ones. In fact, here opinions have a
very general meaning and can represent
the style of life, the way of thinking or of dressing etc.
Then we 
do not consider periodic boundary conditions and 
assume that $x_i \in ] -\infty +\infty[ ~~ \forall i=1,...,N$. Thus
two opinions can  diverge in the time evolution.
Furthermore, it is quite natural that individuals with very
different opinions will tend to interact less.
Such an idea is present in the approach by  Deffuant et al. 
where it is taken into account  by means of the parameter
$\epsilon$.
In our model we assume, instead,
that the coupling between two agents is a decreasing function of
their opinion difference and should vanish
when their opinions are very different.
Finally, as stressed in the introduction,
in our model we are not so much interested in the particular
opinion of each agent: 
rather, we want to see under what conditions it is possible 
to find  agreement among agents with a different velocity ($rate$) 
in changing their opinion. 
Thus, the name of the model: the {\it Opinion Changing Rate}
(OCR) model. 
Taking into account all the previous requirements, we have adopted 
the following dynamics for the OCR model:
\begin{equation}
    \dot{x_i} (t)  = \omega_i + \frac{K}{N} \sum_{j=1}^N
      \sin ( x_j  - x_i ) e^{- \alpha |x_j  - x_i| }  ~~~~~i=1,\dots,N
\label{OCR_eq1}
\end{equation}
Here $x_i (t)$ is the opinion (an unlimited real number) of the $i$th 
individual at time $t$, 
while $\omega_i$ represents the so called
\textit{natural opinion changing rate}, i.e.
the intrinsic inclination, or natural tendency, of each individual
to change his opinion (corresponding to the {\it natural frequency} of each oscillator in the Kuramoto model). 
As in the Kuramoto model, also in the OCR model the $\omega$'s are
randomly drawn from a given symmetric, unimodal  
distribution $g(\omega)$ with a first moment $\omega_0$. 
Usually a uniform or a Gaussian distribution centered at $\omega_0$
is  used. In this way we simulate the fact that in a population there 
are: 1) conservative individuals,
that naturally tend to change their opinion very slowly, and thus are characterized by a value of $\omega_i$ smaller
than $\omega_0$; 2)  more flexible people, with $\omega_i \sim \omega_0$, that change idea more easily
and follow the new fashions and trends;
3) individuals with a value of $\omega_i$ higher than $\omega_0$,
that tend to anticipate the others with new ideas and insights. 
In the equation (\ref{OCR_eq1})
$K$, as usual, is the coupling strength.
The exponential factor in the coupling term
ensures that, for reciprocal distance higher than a certain
threshold, tuned by the parameter $\alpha$,
opinions will no more influence each other, see Fig.2 .
In the following, without loss of generality, we fix $\alpha=3$.
We also notice that, because the natural frequencies $\omega_i$ remain
always constant in time, in absence of interaction (i.e. for K=0) each opinion would change at a constant rate, equal to its associated natural rate, independently from the other opinions.
%
\begin{figure}[htbp] 
\vspace*{8pt}
\centerline{\psfig{figure=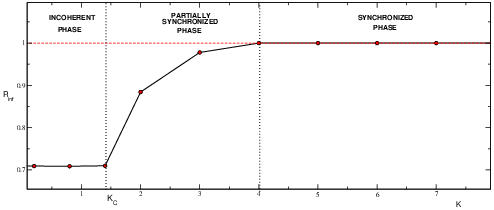,width=12truecm,angle=0}} 
\vspace*{8pt}
\fcaption{Asymptotic order parameter $R_\infty$ as 
a function of the coupling $K$ in the OCR model. Simulations have 
been performed for N=1000 agents and $\alpha =3$ (see text).}
\label{fig2}
\end{figure}
%
\\
The aim of this paper is to study the opinion dynamics of the OCR model
by solving numerically the set of ordinary differential equations
(\ref{OCR_eq1}) for a given distribution of the $\omega$'s 
(natural opinion changing rates) and for a given coupling strenght K.
In particular, we want to find out if, as a function of K, there is
a transition from an incoherent phase, in which people 
tends to have opinions changing at different velocity,  
i.e. at different 'opinion changing rate', to a synchronized 
one in which all the people change opinion with the same velocity. 
The latter, in this case, will represent the common trend of the
entire society, or, in other words, the common rate of the
'\textit{changing world}'.
\\
Because of the non-periodicity of the opinion values in our model, 
we cannot use the Kuramoto mean field order parameter 
of Eq.(\ref{kuramoto_eq2}) to measure the degree of synchronization.
On the other hand there are a series of different alternatives. Here we adopt an
order parameter related to the standard deviation of the
opinion changing rate $\dot{x_j}(t)$.
Such a parameter, indicated with $R(t)$, is defined as:
\begin{equation}
R(t) = 1 - \sqrt{ \frac{1}{N} \sum_{j=1}^N (\dot{x}_j(t) - \dot{X}(t))^{2}}
\label{OCR_orderpar}
\end{equation}
where $\dot{X}(t)$ is the average over all individuals of $\dot{x_j}(t)$.
>From Eq.(\ref{OCR_orderpar}) it follows that:

\begin{itemize}
\item
$R=1$ in the fully synchronized phase,
where all the agents have exactly the same opinion changing rate
(and very similar opinions).
\item
$R<1$ in the incoherent or partially synchronized phase,
in which the agents have different opinion changing rates and
different opinions.
\end{itemize}
\subsection{Phase Transition}
\noindent
Our numerical simulations have been performed typically
with N=1000 agents and with an uniform distribution of the
initial individual opinions $x_i(t=0)$ in the range [$-\Delta,\Delta$].
Such a range has to be chosen consistently with the parameter
$\alpha$ in the coupling term (see Fig.2), in order to ensure that at the
beginning of the simulation all the agents would interact with each
others. In the following we will set $\Delta = 1$.
The natural opinion changing rates $\omega_i$ are
taken from a uniform distribution in the range [0,1] 
\footnote{As for the Kuramoto model, in the OCR model a uniform 
distribution of opinion changing rates in the range 
[0,$\omega_{max}$] can always be trasformed into the range [0,1] 
by a scaling of the coupling constant $K^{\prime} = K \cdot \omega_{max}$.}.
We have checked that the alternative choice of a Gaussian
distribution does not change significantly the results
of the simulations.
We fix the value of the coupling $K$ and we let
the system evolve by means of Eq.(\ref{OCR_eq1})
until a stationary value for the order parameter $R_\infty = R(t=\infty)$
is obtained.

\begin{figure}
\centerline{\psfig{figure=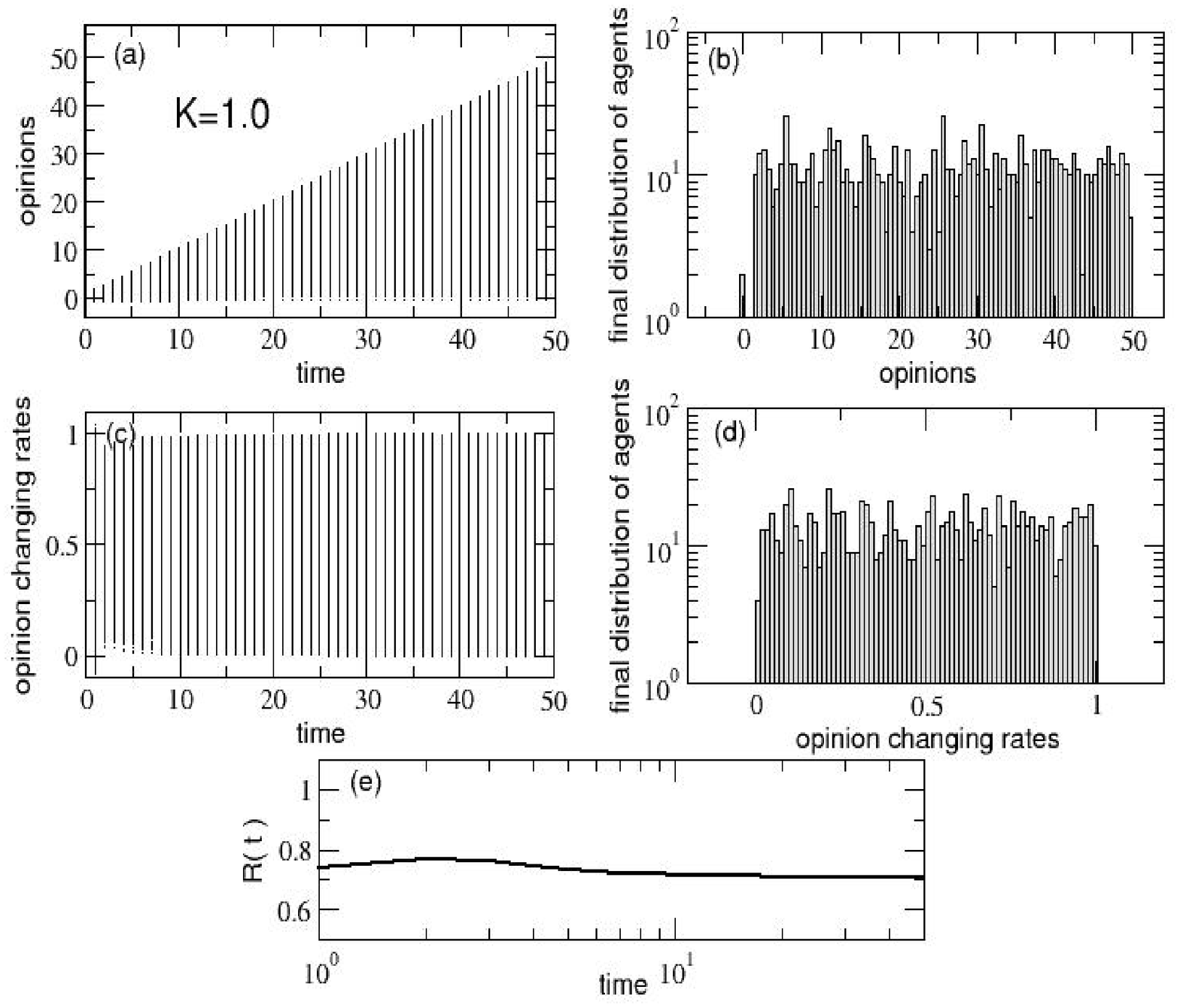,width=12truecm,angle=0}}
\vspace*{8pt}
\fcaption{OCR model for $N=1000$ and $K=1.0$. In
panels (a) and (c) we plot the time evolution of the opinions
and of the opinion changing rates, while in panels (b) and (d) we show 
the respective final distributions of agents. 
Notice that, in panels (a) and (c) 
we report, for each time step, 1000
points corresponding respectively to 
the opinions and opinion changing rates of the N agents.
In panel (e), 
we report the time evolution of the order parameter $R$. 
Being in the incoherent phase, 
the system rapidly reaches an homogenous state 
for the opinions as well as for the frequencies,
with a saturation value of the order parameter less than one.}
\end{figure}
%
\begin{figure}
\vspace*{8pt}
\centerline{\psfig{figure=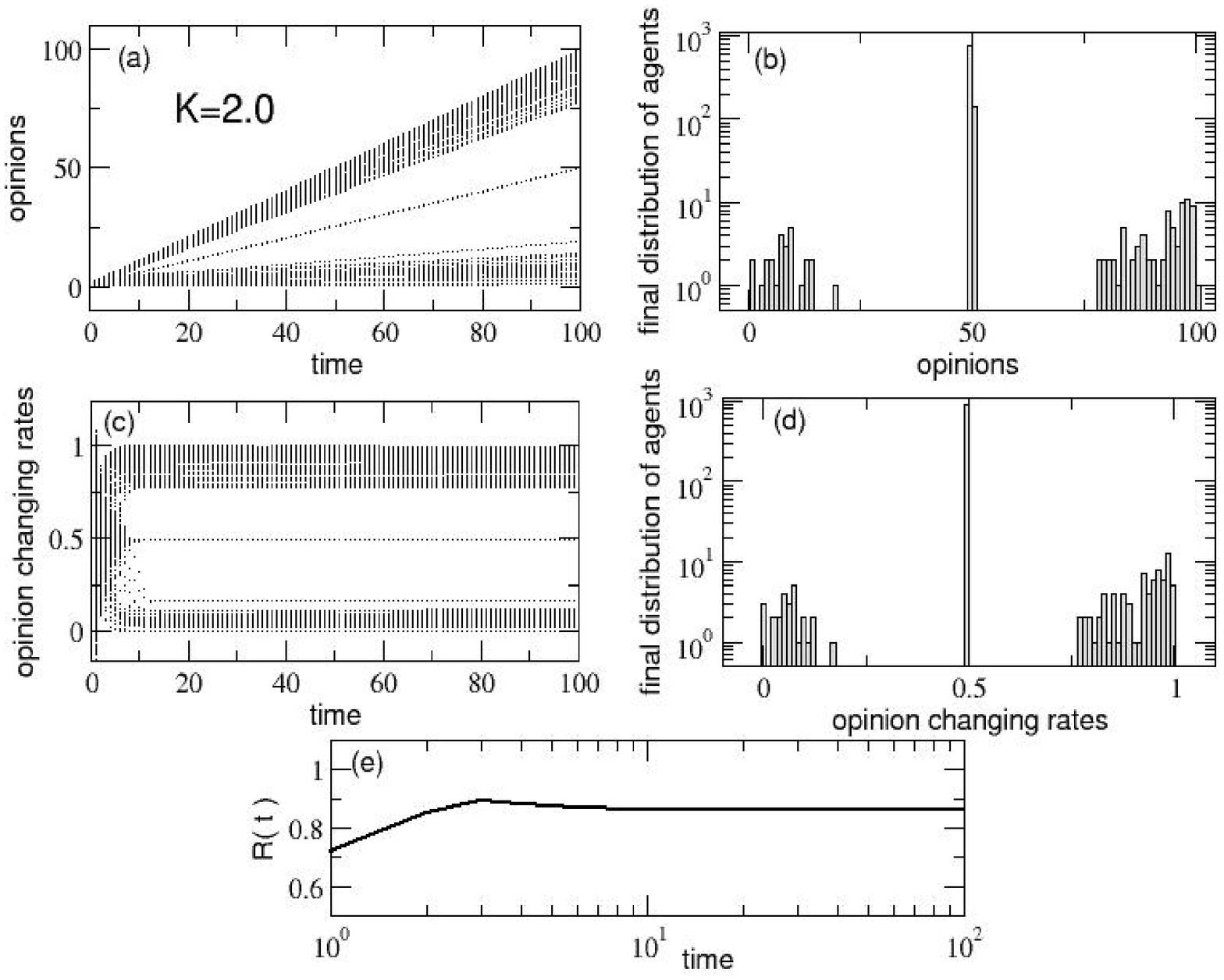,width=12truecm,angle=0}} 
\vspace*{8pt}
\fcaption{OCR model for $N=1000$ and $K=2.0$. Same quantities as in Fig.4.
In this case we are in the partially synchronized phase and the system 
splits into three clusters: a central one, with the largest part 
of the agents, and two lateral ones. 
The saturation value of the order parameter is still less than one, 
but it is larger than in the incoherent phase (previous figure).}
\end{figure}
%
\begin{figure}
\vspace*{8pt}
\centerline{\psfig{figure=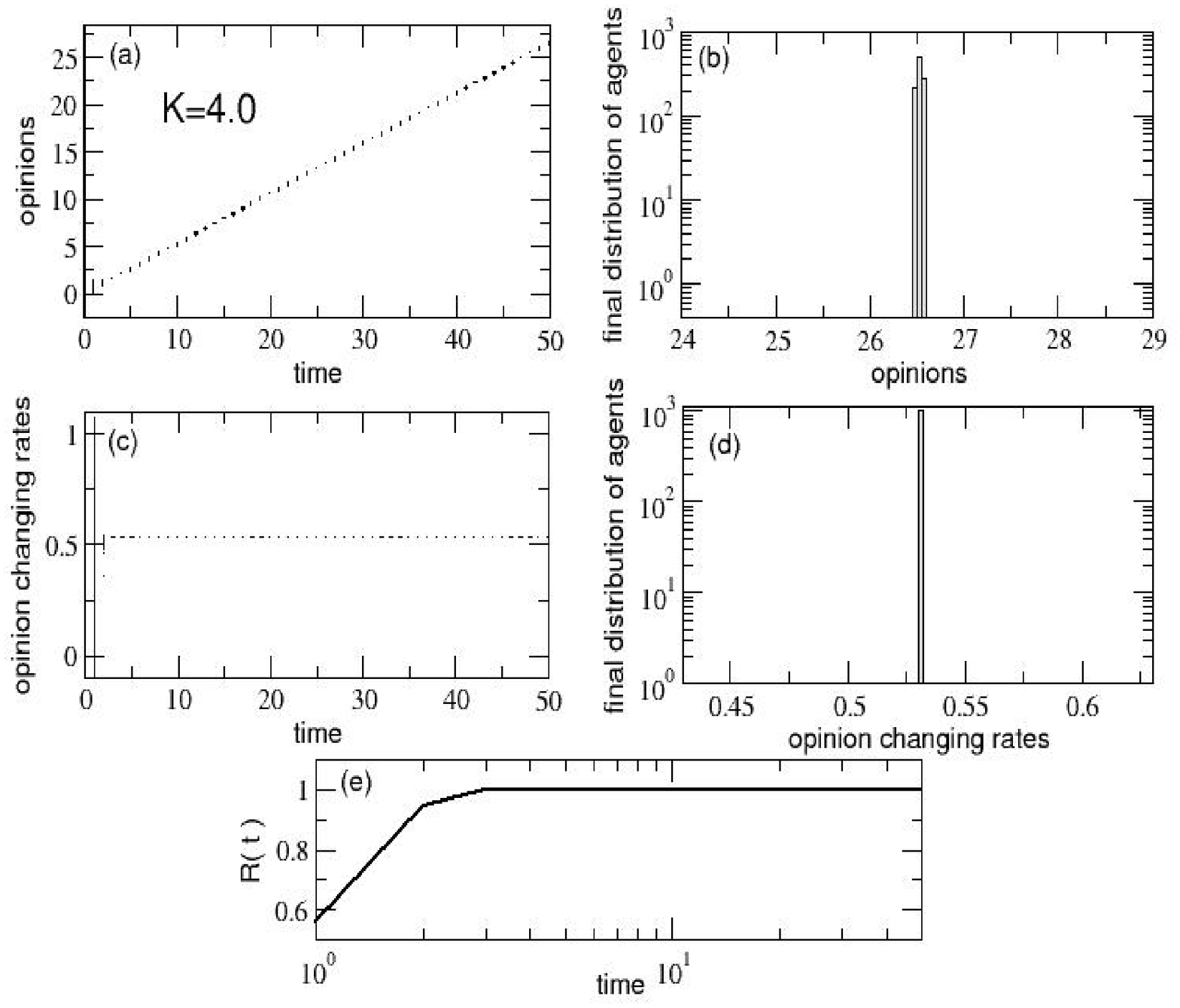,width=12truecm,angle=0}} 
\vspace*{8pt}
\fcaption{OCR model for $N=1000$ and $K=4.0$. 
Same quantities as in Fig.4. In this case the coupling is strong enough 
so that all the agents reach a synchronized state (frequency locking). 
We observe a single final cluster, both in opinions and in opinion 
changing rates. The  asymptotic order parameter rapidly becomes 
equal to one.}
\end{figure}
%
\begin{figure}
\vspace*{8pt}
\centerline{\psfig{figure=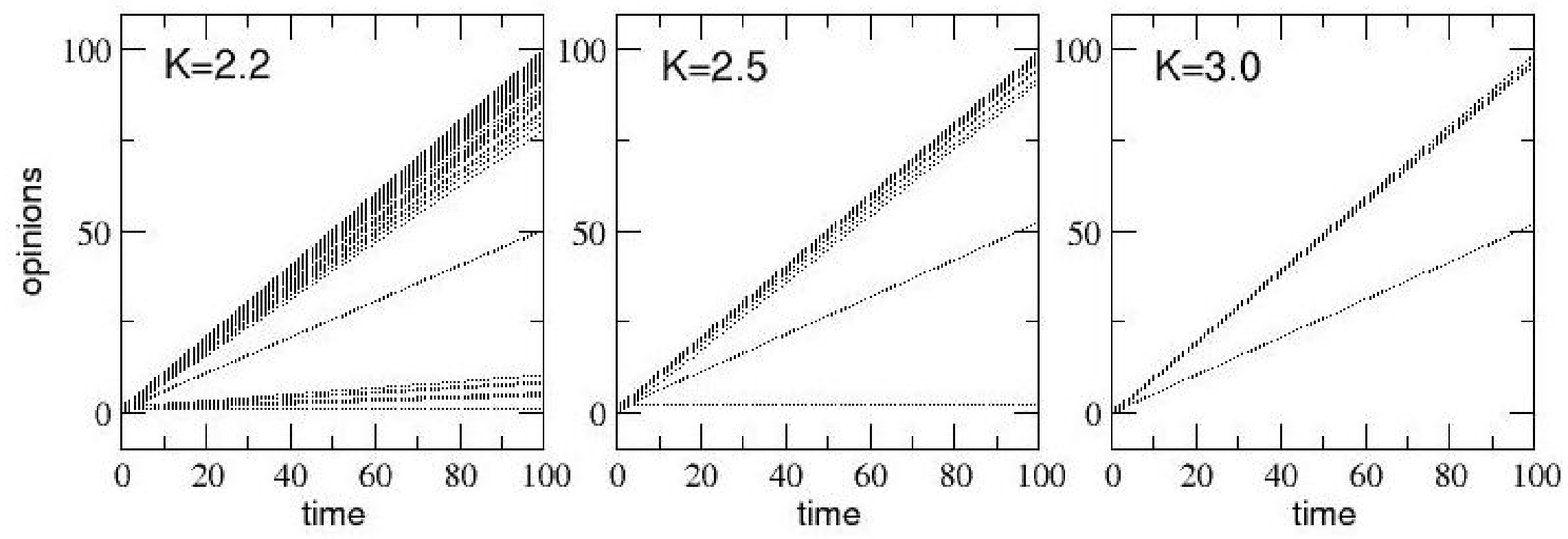,width=12truecm,angle=0}} 
\vspace*{8pt}
\fcaption{Time evolution of opinions for three values of the coupling $K$
in the partially synchronized phase. For values of $K$ larger than 2.0 
(compare with Fig.5), the 'conservative' group rapidly vanishes.
The same thing happens to the 'progressist' group for $K$ larger than 3.0}
\end{figure}
%
%
In Fig.3 we plot the asymptotic order parameter as a function
of $K$ for a system with $N=1000$ agents.
It is easy to recognize a Kuramoto-like transition from an
incoherent phase (for $K<K_c\sim1.4$) to a partially coherent
(for $K\in[1.4,4.0]$) and, finally, to a fully synchronized
phase (for $K>4.0$).
A better characterization of $K_c$ as a function of N, $\alpha$ and $\Delta$
is under current study.
We now focus on the details of the dynamical evolution in each of
the three phases.
\\
In Fig.4 we show the time evolution of the opinions and of the
opinion changing rates (angular velocities or frequencies)
with the respective final distributions for a small value of
the coupling, namely $K=1.0$.
On the bottom part of the figure we plot the order parameter time
evolution.
Being in the incoherent phase, the system remains in a
configuration of agents with finite standard deviation 
of opinion changing rates: because of the weak interaction 
with the others, each agent keeps his natural 
opinion changing rate and the different opinions 
will diverge in time without reaching any 
kind of consensus. We could also look at this 
as to an $'anarchical'$ society. In such a phase the order 
parameter R takes the minimum possible 
value that, at variance with the Kuramoto's model,
here is not zero.
\\
In Fig.5 we show the case $K=2.0$. The coupling is still weak but strong
enough to give rise to three different clusters of evolving opinions,
each with a characteristic changing rate: the largest number of the agents,
representing "the world's way of thinking" or, if you want,
the "public opinion", moves with an intermediate angular velocity,
but there is a consistent group of people remaining behind
them and also a group of innovative people (quicker in supply new
ideas and ingenuity). From a political point of view, we could interpret
this situation as a 'bipolarism' with a large number of 'centrists'.
In this case the order parameter is larger  
than in the previous example, but still less than one because 
the opinion synchronization is only partial.
\\
Finally, in Fig.6 we report the case $K=4.0$.  The coupling is so
strong that all the opinions change at the same rate: in the histograms  
we observe a single final cluster, both in opinions and in opinion  
changing rates. It is important to stress, however, 
that, as in the Kuramoto model \cite{sebino} also in the 
equations (\ref{OCR_eq1}) of the OCR model 
is impossible to synchronize perfectly both phase and frequency. 
In fact, all the oscillators assume \textit{exactly} 
the same frequency (frequency locking) and \textit{almost} 
the same phase, with a small phase shift between 
them \cite{sebino}. 
In this $'dictatorial'$ society all the agents
think in the same way and follow the same trends 
and fashions. Although the $natural$ frequencies 
of the agents are - as in the previous examples -
different from each others, their opinion changing rates rapidly
synchronize (frequency locking) and thus the order parameter R 
reaches a saturation value equal to one. 
\\
It is interesting to notice that, by increasing the value of $K$ from 2.0 
to 4.0 (see Fig.7), the group with a low opinion changing rate 
rapidly disappears, leaving place only to two groups, 
the central one and the innovative (fast) one. 
The survival of the fast group is not fully understood although it 
seems to support the Darwin's statement  reported in the introduction.

Above $K=3.0$ also the fast group vanishes until, near $K=4.0$, 
only the central synchronized group remains. 
Some conclusions, in agreement with the intuition, can be drawn 
from the numerical simulations we have performed. 
In the OCR model, 
in order to ensure a 'bipolarism', i.e. 
an equilibrium between the conservative and progressist 
components, a changing society needs a level of coupling $K$
strictly included in a narrow window ($1.5 < K < 2.5$) inside 
the partially synchronized phase. 
Otherwise such an equilibrium will be broken and the final result 
will be anarchy or dictatorial regime.

\subsection{Metastability}
\noindent
In the previous subsection we have shown under what conditions the OCR model
exhibits dynamical synchronization in the agents' frequency and opinion.
In this subsection we investigate the dynamics of the OCR model
in the case in which the initial opinions are synchronized from the beginning. Instead
of extracting the initial opinions 
from a uniform distribution with halfwidth $\Delta =1$, as in the previous simulations, here we start all the agents with the same
opinion ($\Delta=0$).
The distribution of the natural inclinations to change, $g(\omega)$, is the same
as considered in subsection 3.1.
%
\begin{figure}[htbp] 
\vspace*{8pt}
\centerline{\psfig{figure=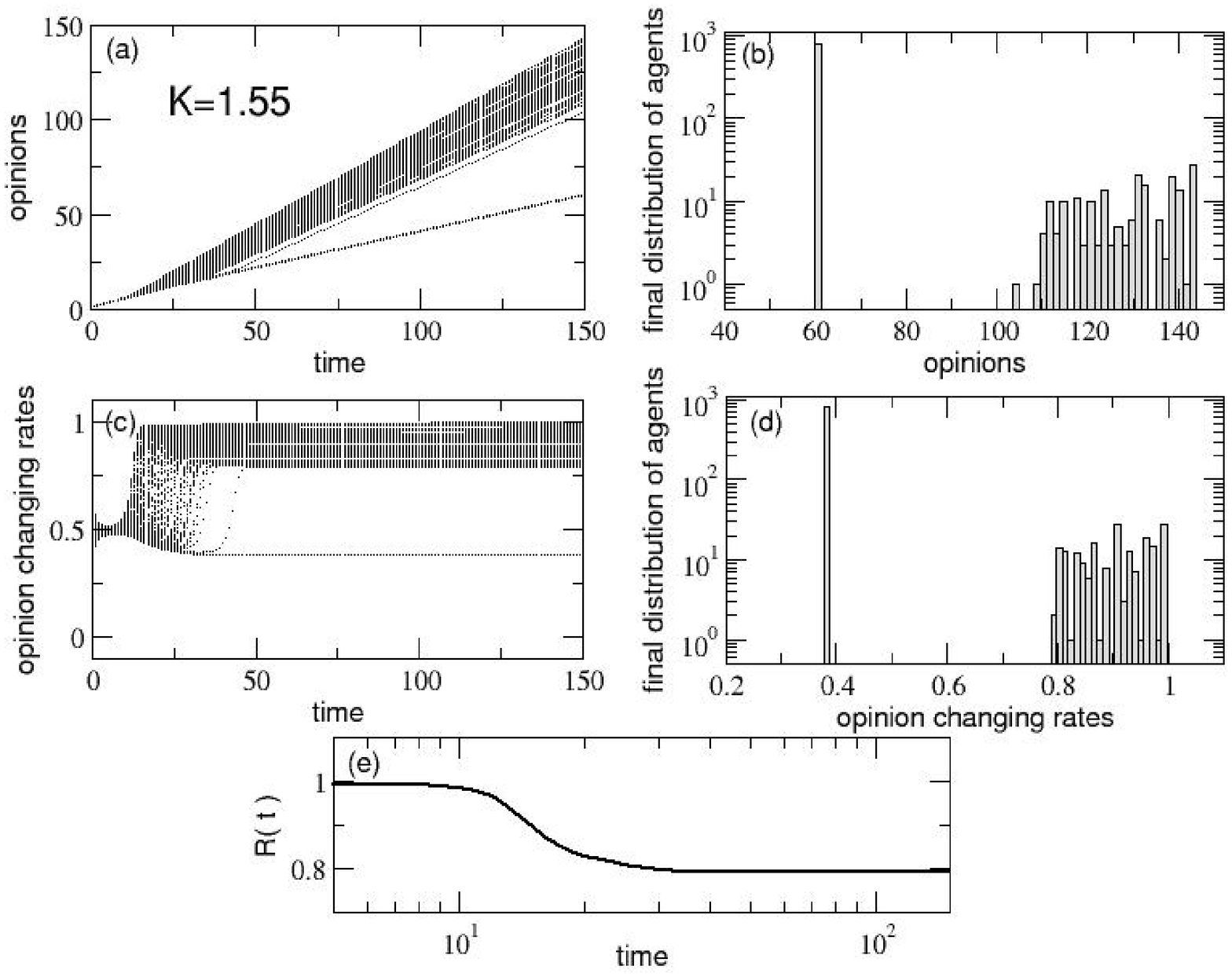,width=12truecm,angle=0}} 
\vspace*{8pt}
\fcaption{OCR model for N=1000, K=1.55
and completely synchronized initial opinions.
The system remains synchronized for a short time and then
relaxes to a stationary state where,
apart from the main synchronized group,
a largely spread cluster of progressists appears.}
\end{figure}
%
In Fig.8 we report the results of the numerical simulation for
N=1000 and $K=1.55$, a value of the coupling constant
higher than the critical value $K_c=1.4$.
We observe that the opinion changing rates initially try to synchronize
in only one big cluster, and all the agents have very similar
(but not identical) opinions. Then, rapidly and progressively, many
agents leave the (almost) common opinion and increase
their opinion changing rates, concentrating in a new,
largely spread, cluster of progressists. Also in this case,
the conservative group observed in Fig.5 does not appear.
Correspondingly, the order parameter R stays for a while
in a metastable state around the value $R=1$, 
then rapidly decreases and stabilizes at $R=0.8$.
%
\begin{figure}[htbp] 
\vspace*{8pt}
\centerline{\psfig{figure=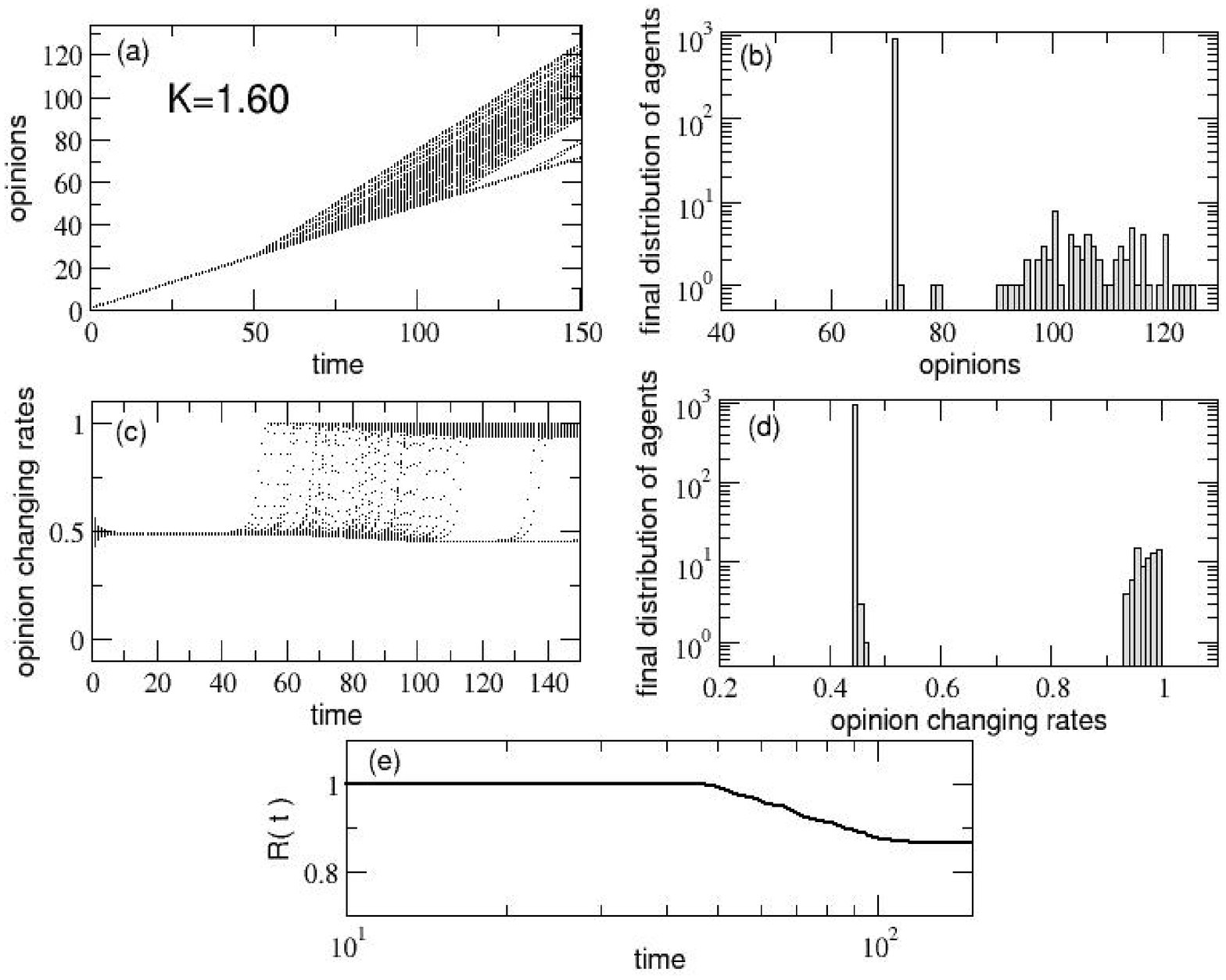,width=12truecm,angle=0}} 
\vspace*{8pt}
\fcaption{OCR model for $N=1000$, $K=1.6$
and completely synchronized initial opinions. Same plot as in Fig.8
The initial synchronized state behaves 
as a metastable state whose life-time diverge 
approaching the value of the coupling $K\sim1.62$}
\end{figure}
\\
In Fig.9 we report the results obtained for a larger value of K,
namely $K=1.6$. Here the life-time of the metastable state
with $R=1$ becomes visibly longer. We have checked that,
approaching $K\sim1.62$, the life-time of such a metastable state
diverges exponentially. 
The presence of metastable states with a diverging life-time
for values of the control parameter above the transition
from the homogeneous to the non-homogeneous phase is a characteristics
observed in a large class of systems \cite{metastable}. It seems to 
be related to a glassy-like dynamical behavior that would hinder
synchronization \cite{glassy,kuraglassy}.
The social consequence of this out-of-equilibrium scenario could be the metastability of a dictatorial regime (chosen as initial condition) approaching a critical value of interconnections between individuals. 
%
\begin{figure}[htbp] 
\vspace*{8pt}
\centerline{\psfig{figure=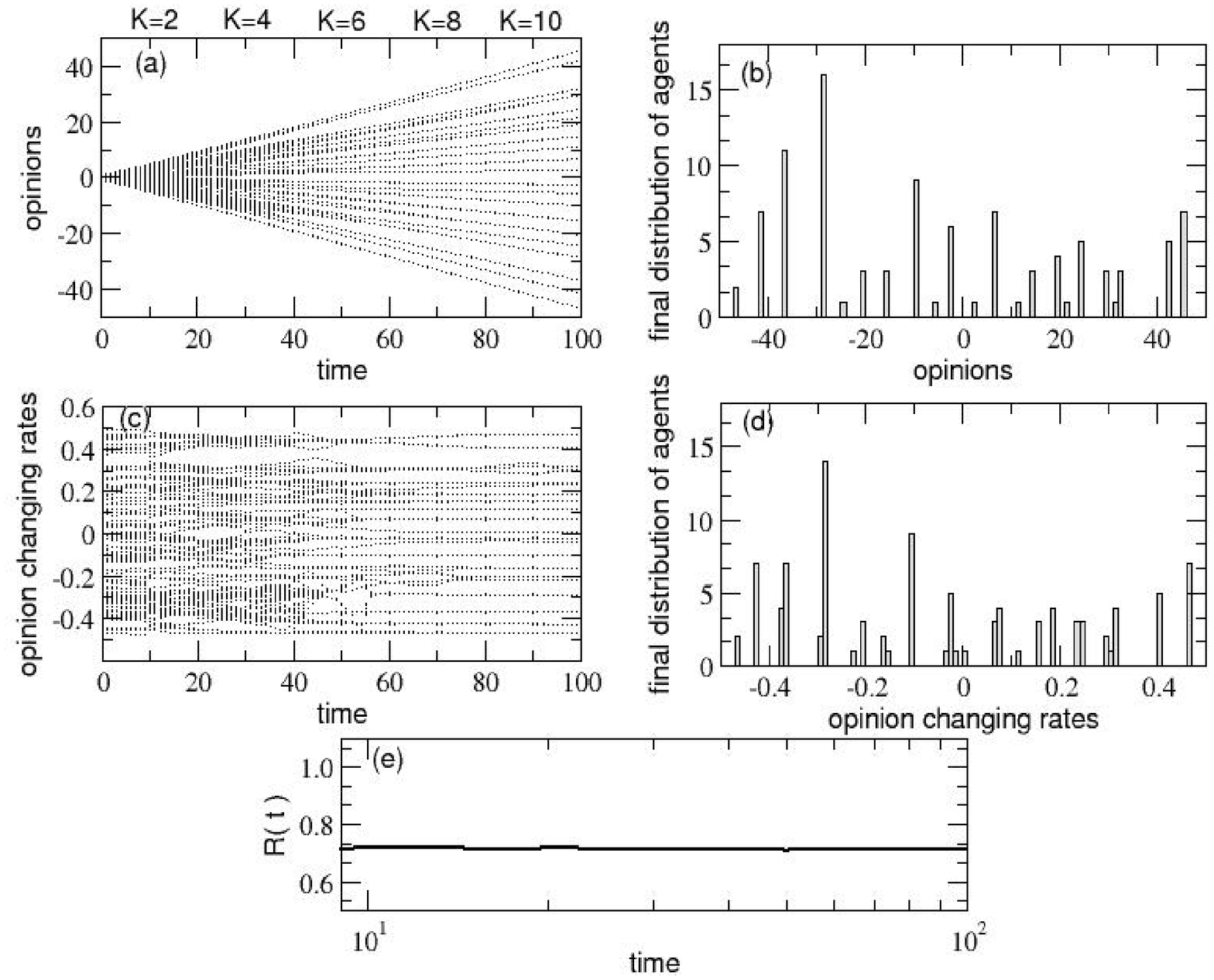,width=12truecm,angle=0}} 
\vspace*{8pt}
\fcaption{OCR model for $N=100$ and a value of 
$K$ increasing with a constant rate from 0.1  
up to 10.1. We plot the same quantities as in the previous figures}
\end{figure}

\subsection{Variable Coupling}
\noindent
Finally we want to see what happens if the coupling
$K$ is let, respectively, to increase or decrease its value 
during the dynamics. In this way we can simulate 
a society in which the interconnections between the agents 
changes in time. 
The initial opinions and the natural changing rates are, 
as usual, uniformly distributed. At variance with the previous 
simulations, and only for a better visualization of the outputs,   
here we consider $N=100$ and the natural opinion changing 
rates distributed in the interval $[-0.5,0.5]$.   
\\
In Fig.10 we consider a case in which the coupling is 
uniformly increased from $K=0.1$ to $K=10.1$.  
The agents' opinions initially spread freely, 
and then rapidly freeze in a large number of non-interacting 
clusters with different changing rates and variable sizes. 
Note that the final distribution of clusters is different 
from those in Fig.4, even if also in this case 
the order parameter R reaches the same minimum value, 
since the non-interacting clusters are uniformly distributed. 
The particular cluster distribution observed in Fig.10 
cannot be obtained in simulations with a constant coupling. 
This could suggest that the increase of interactions between 
the members of a society (due to the improvement in transport 
and communications) would be determinant in order 
to stabilize that plurality of different (and consistent) 
clusters of opinions (ideologies, political parties, etc.) 
necessary for a democracy.
\\
\begin{figure}[htbp] 
\vspace*{8pt}
\centerline{\psfig{figure=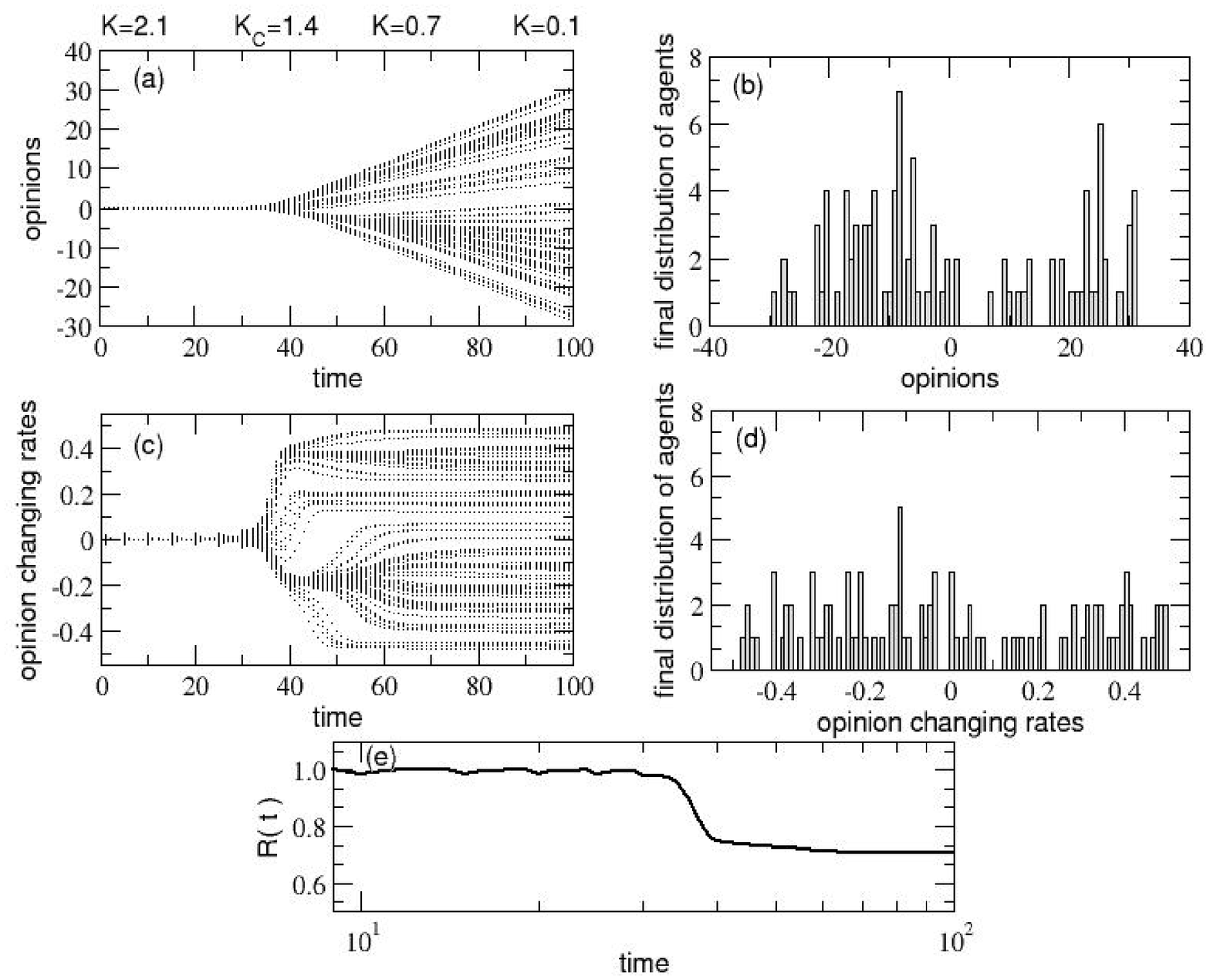,width=12truecm,angle=0}} 
\vspace*{8pt}
\fcaption{OCR model for $N=100$ and a value of 
$K$ decreasing with a constant rate from 2.1 to 0.1. 
We plot the same quantities as in the previous figures.}
\end{figure}

In Fig.11, we consider the case in which the coupling is uniformly 
decreased from $K=2.1$ (a value in the partially synchronized phase) 
to $K=0.1$ (incoherent phase). 
Also in this case, after a synchronized initial evolution, 
as soon as the coupling exceed the value $K_C$, 
the opinions spread.  
However  they do not form the same clusters as in the  previous simulation:  
they behave more incoherently showing a slight 
tendency to form two groups with opposite opinion changing rates. 
This is  typical of the partially synchronized case from 
which the system started (see Fig.5). Finally, 
the order parameter, from the unitary value corresponding 
to the synchronized initial evolution,  
goes down to about the same value of the previous case (Fig.10). 
This scenario  reminds the typical situation of the fall of a dictatorship, 
which is usually followed by a chaotic regime (see for example the situation 
in Iraq after the Saddam Hussein capture). In alternative, it could be 
interpreted as the dissolution of an empire in many little countries 
(think of the dissolution of the Roman Empire or, more recently, to 
the dissolution of the Soviet Union).

\section{Conclusions}
In spite of its simplicity, the OCR model seems to capture 
some general features of the opinion formation process. The model is based 
on a set of coupled ordinary differential equations governing the rate of 
change of agents' opinions.  
In particular with the OCR model we have introduced the concept 
of \textit{'opinion changing rate'}, 
transforming the usual approach to opinion consensus 
modeling \cite{Sznajd,Deff,HK} into a synchronization problem: 
``how can one synchronize the opinion of many agents having a different 
natural inclination to change idea?''. 
The OCR model shows many interesting features with a given 
social meaning. 
First of all, the model exhibits a phase transition from 
an incoherent phase to a synchronized one at a well defined critical 
threshold $K_c$. 
We have observed three main scenarios as a function of the 
coupling $K$, which can be interpreted as the degree of 
interconnection and/or communication among individuals.  
\\
The incoherent state emerging for $K<K_c$ can be interpreted, in social terms,  
as anarchy in a society of many individuals with scarce or disrupted communications. 
In alternative, if we generalize the concept of agents to represent 
groups of individuals, for instance social communities, instead of single individuals,    
the state we obtain for $K<K_c$ can be interpreted as a world with many 
isolated and non interacting cultures. 
On the other hand, when $K \gg K_c$, i.e. in the fully synchronized phase, 
the model shows the appearance of a dictatorial regime,  
or a globalized world, where social and cultural differences are  
constrained into a single way of thinking, notwithstanding the 
different tendencies of each individual. 
Finally, we found that bipolarism and democracy are possible only in an 
intermediate regime, with $K \sim K_c$, 
corresponding to a partially synchronized phase. 
In the latter regime the role of the initial conditions and of the dynamics  
becomes particularly important. For instance, the model shows,  
as we approach a critical value of interconnections among individuals, 
the metastability of a dictatorial regime, chosen as initial condition. 
\\
We have also simulated various situations in which the coupling $K$ is changing in time. 
In such cases we have observed a clustering of opinions 
(for $K$ increasing in time) and the breaking of sinchronization 
(for $K$ decreasing in time), respectively corresponding to the coexistence 
of many political parties, and to the dissolution of an empire or a 
dictatorship.   
\\
Future developments include more analytical approaches to the OCR model 
and some generalizations in order to study opinion synchronization 
in more complex situations, as for instance: 

\begin{itemize}

\item social agents placed on a complex network \cite{lmprl,lmepjb}.  
In this way one can investigate the influence of the topology on 
the opinion formation dynamics, with particular attention to 
the role of node centrality \cite{lmcentral}. 

\item the introduction of an external field representing the 
pressure of mass-media. 

\item the addition of disorder in the coupling among the agents. 
\end{itemize}

\nonumsection{Acknowledgements}
\noindent
We thank Santo Fortunato for having introduced us to the 
the fascinating field of sociophysics.

\nonumsection{References}

\end{document}